# A Cybersecurity Guide for Using Fitness Devices

Maria Bada[1 2][0000-0003-0741-5199] and Basie von Solms[2][0000-0003-3586-6632]

[1]University of Cambridge, Cambridge CB3 0FD, UK
`maria.bada@cl.cam.ac.uk`

[2] University of Johannesburg, Auckland Park, Johannesburg, South Africa
`basievs@uj.ac.za`

**Abstract.** The popularity of wearable devices is growing exponentially, with consumers using these for a variety of services. Fitness devices are currently offering new services such as shopping or buying train tickets using contactless payment. In addition, fitness devices are collecting a number of personal information such as body temperature, pulse rate, food habits and body weight, steps-distance travelled, calories burned and sleep stage. Although these devices can offer convenience to consumers, more and more reports are warning of the cybersecurity risks of such devices, and the possibilities for such devices to be hacked and used as springboards to other systems. Due to their wireless transmissions, these devices can potentially be vulnerable to a malicious attack allowing the data collected to be exposed. The vulnerabilities of these devices stem from lack of authentication, disadvantages of Bluetooth connections, location tracking as well as third party vulnerabilities. Guidelines do exist for securing such devices, but most of such guidance is directed towards device manufacturers or IoT providers, while consumers are often unaware of potential risks. The aim of this paper is to provide cybersecurity guidelines for users in order to take measures to avoid risks when using fitness devices.

## 1 Introduction

Currently there is a lot of discussion around the Internet of Things (IoTs) and the new types of threats emerging from new technologies [1] [2]. IoTs create great opportunities for everyone but also create new tools for citizens and criminals.

Wearable devices are becoming very popular especially in healthcare and fitness. Such devices could be health monitors, fitness bands or smartwatches. In addition, the worldwide sales of these devices have been growing with 110 million units in 2018 [3].

Wearable devices are quite popular, since they merge the physical and digital world. A recent study [4] identified that smartphone users tend to also own wearables. Moreover, a large percentage of users would use wearables in the next five years not only to track health related information, but also for everyday actions such as unlocking a door, making financial transactions or authenticating their identity.

Some devices such as Fitbits are currently offering new services such as making transactions with contactless payment for example in order to do shopping or buy train tickets. In order to use the device for such services consumers need to add their credit or debit card to their Fitbit watch or tracker [5]. Fitbit Pay works with a number of Fitbit smartwatches [6]. This function is also supported by the consumer's bank fraud protection services. In the UK, the device can be used for Transport for London (TfL) transit system including buses, London Underground and trains and other services [5].

However, more and more reports are warning of the cybersecurity risks of such devices, and the possibilities for such devices to be hacked. Guidelines do exist for securing such devices, but most of such guidance is directed towards device manufacturers, IoT providers and more. One good example is the Code of Practice for consumer IoT security by the UK Department of Culture Media and Sport [7]. Very little, if any such guidance does exist for the real end user, the ordinary citizen, who makes use of fitness devices and other smart home devices.

The purpose of this paper is to put the focus on the end users and emphasise on the security risks of fitness devices. The focus of this research lies on fitness devices due to their wide current usage and the direct or indirect risks they pose to users. Risks are not only related to security but also to user privacy. In order to create a set of guidelines for users around security and privacy issues, we reviewed current research following a grounded theory approach.

In Section 2, we discuss the personal data collected by fitness devices while in Section 3 we provide a description of the cybersecurity risks they are exposed to. In Section 4 we discuss the cybersecurity awareness needs of users of wearable devices and in Section 5 we present a number of guidelines for users and manufacturers in order to lower their exposure to risks. Finally, Section 6 discusses the findings and Section 7 concludes this paper.

## 2    Personal Data Collection by Fitness Devices

Different models of wearable devices are offered to consumers from different companies. For example, Fitbits, Apple Watches and Samsung Galaxy Watches are some of the products available today. These devices have a particular function; however, they also collect personal data from users [8]. We will review Fitbit as an example, however other products collect similar information. For example, Fitbit smartwatches collect information that can be considered private and potentially dangerous in the wrong hands. According to Fitbit's Legal Policy, Fitbit receive or collect three categories of information from user devices [9]:

- *Physical Data:* data such as body temperature, pulse rate, food habits and body weight, steps-distance travelled, calories burned, sleep stage, and active minute. These data are synchronised to devices that transfer information to Fitbit servers.
- *Location Data:* Fitbits receive information regarding location through GPS signals or Wi-Fi access points.
- *Usage Data:* Fitbits collect information about a user's interaction with services, such as when a user views or searches content or installs software.

Users typically transmit this data using a wireless connection, such as a Bluetooth, Wi-Fi or a cellular connection. For example, devices transfer the data they collect through a smartphone using Wi-Fi or Bluetooth connections. The data received are then stored or transmitted to a cloud server [10].

Fitbit introduced a series of technology for workout tracking such as PurePulse, SmartTrack and Sleep Tracking a technology that automatically recognizes users' exercises and record the data through a smartphone [11].

## 3 Risks Arising from Fitness Devices

Although fitness devices have been developed and used for a number of years, it seems that developers still do not focus on the security of these devices. As mentioned above, these devices collect information related to the health and wellness of consumers, thus collecting a big amount of data. As a result, device manufacturers are often putting consumer data at risk of exposure due to device related vulnerabilities.

The main types of security incidents related to Fitbits are based on [9]:

1. *Lack of Authentication and Physical Security Control:* Currently, Fitbit smartwatches do not have a built-in security mechanism. Without authentication, Fitbits pose a threat to an individual's personal information. Due to the lack of authentication these devices are placed as highly vulnerable, since a hacker can gain access to a company's network through these for potential exploitation.
2. *Disadvantages of Bluetooth Connections:* Since the main connectivity method for a Fitbit is Bluetooth, it inherits the same vulnerabilities that most Bluetooth devices have during times of communication. Fitbits are susceptible to various threats such as message modification, denial of service (DDoS), and eavesdropping attacks.
3. *Location/Tracking and Biometric Leakage:* A Fitbit can acquire information regarding a user's location, either through the built-in GPS, or by pairing with a smartphone's GPS. Location tracking can raise serious security concerns for individuals and organisations.
4. *Third-Party Related Attacks*: In 2015, a big number of online accounts of Fitbit users were attacked by hackers. The hackers used details such as email addresses and passwords from third-party websites to log in to Fitbit accounts. They then used the details to file false claims for replacement orders. Also, they managed to gain access to customer personal data such as GPS history [12].

As mentioned above, wearable devices allow the collection of personal data through device sensors initially while the data are then shared through a smartphone [13]. Due to their wireless transmissions, these devices can potentially have vulnerabilities to a malicious attack allowing the data to be exposed. Because a wearable, like a Fitbit exercise watch, is mostly connected and synchronised with the user's smart phone or laptop, infecting the Fitbit with malware will automatically infect the user's smart phone when fitness data is uploaded. Once the phone is infected, all possible compromises can happen, and the infection can be then propagated to other trackers [14].

For example, hackers could wirelessly upload malware onto a Fitbit by using Bluetooth. Although a hacker would need to be near the targeted Fitbit in order to infect it, the Bluetooth connection might take place in any public area such as in a park or coffee shop. The process would take only 10 seconds, and the user would not notice anything wrong with their Fitbit at that point. Once the user connects the infected wearable device to a PC or laptop, the malware can spread to the personal computer or the business computers and even the entire network [15].

In order for information to be transmitted between two devices, one device must establish a central role in the connection with Bluetooth and the second device must play a peripheral role [11]. For example, in the case of a pair of Bluetooth Fitbit SmartTrack to iPhone, the iPhone would play the role of the central device and Fitbit SmartTrack would be the peripheral device that indicates available connection where the signals contain the IP address of a mobile device and a payload containing data about the connection [11].

Also, these devices include tools such as accelerometers and gyroscopes, which provide useful data with which someone can identify fine motor task movement [16], and orientation [17]. If the potential risks are considered a step further, then it is clear that such devices can be used for more sophisticated cyber-attacks by capturing the gentle movements and position changes of the wrist while writing.

Such sophisticated tools can be used to recognise the password typed by a user or the security number of an employee when they type it in real time. Using machine learning, researchers were able to detect the wrist movement while writing digits and were able to construct a robust machine learning model predicting perfect real-time performance [18].

Other security related risks are arising from the payment facilities, through the credit card information which can be stored on the user's device. As mentioned above, some devices such as Fitbit are currently offering the possibility to connect the device with a credit card and use it as an electronic wallet. The main reference around security of this functionality is around fraud and the responsibility of the consumer's bank to provide support [5].

## 4     Cybersecurity Awareness of Users

Although users might expect a USB stick to be a way of transmitting malware or become a target for hackers, most users do not expect their fitness trackers to be a target [19]. However, these little devices are the perfect delivery system for malware. People perceive wearable devices as a new mean to interact with their social groups and not as a potential threat to their medical information. In addition, younger people are more likely to share their wearable device's data online [20]. It is often the case that users are willing to sacrifice security and privacy because of the convenience a smart device provides [21].

Users often lack awareness about the information security and privacy related issues of using wearable devices [3]. In addition, users might not know what types of data are being collected or being stored or transmitted by their wearable devices. Moreover, there is a general lack of awareness around encryption of data during transmission. It is therefore, not a surprise that lack of awareness also leads to users not being aware of the security policy for their wearable device or the security measures used to protect their data. In addition, if users suspect an information security incident, they are not aware where to report it.

This lack of awareness of users around privacy and the blind trust of users that their data would be protected might complicate the accountability of service providers in the case of a data breaches [22] [23] [3].

Additionally, there is the issue of ownership of the data [24]. Previous research showed that only a small number of users backed up sensitive or critical data at a regular basis or tested recovery [3]. Therefore, it is obvious that security and privacy considerations of such delicate data collected by IoT devices are essential.

As technology developers push new wearable devices to the market and make these devices sync with existing smartphones, malware creators are looking for new avenues of attack [25]. Criminals can use the data to organise cyber-attacks based on identity theft, impersonation by creating a fake user profile or by using more targeted phishing attacks. Voice-recognition technology might also facilitate such cyber-attacks through the different tools already being used in vishing.

It is important to notice that wearable devices are the first category of information technology (IT) devices where there is not only danger due to the exposure of consumer data, but also the real potential to cause physical harm to wearers. Therefore, it is essential that all users of fitness devices are informed about the potential risks stemming from using them.

## 5      Cybersecurity Guidelines for Users and Manufacturers

Policy makers, but also businesses, NGOs or users, face enormous challenges since current policies are often inadequate [26]. Currently emerging technologies require a new thinking around privacy, freedom of expression, intellectual property protection and national security.

However, policies are not the only necessary measure that needs to be taken. It is important to establish a cybersecurity ecosystem in all sectors and focus more on the attitudes, beliefs and practices of end-users. Dutton [27] describes a cybersecurity mindset as "*a pattern of attitudes, beliefs and values that motivate individuals to continually act in ways to secure themselves and their network of users*".

Consumers need to take some basic steps in order to ensure that their personal information is not exposed to malicious intent. Some of the simplest steps is to research the security features of the intended device and read relevant reviews of the product [28].

However, there is a number of steps users and manufacturers can take in order to lower their exposure to risks. These are summarized in Table 1.

### 5.1 Cybersecurity Guidelines for Users

A number of guidelines which users can follow to ensure a safe use of wearable devices are presented below:

**Adjusting the Settings.** To prevent malware infection, it is important that all personal and business devices are always protected from outside threats. Many devices, such as smartphones or laptops, allow to secure their Bluetooth connection with a password to prevent unauthorized access. This way, Fitbit will only connect to the customer's phone.

Users need to always exercise caution when plugging any device into a computer. Using an antivirus program to scan any connected device for viruses and malware and determining the settings to limit the personal data captured by the device are important steps towards security. Disabling the permission in capturing data such as sleep patterns and capture only the data a user really needs is also a good measure.

**Education on Risks.** A significantly stronger security posture can be achieved simply by educating Fitbit users about keeping their devices and software current with the latest security updates, and the best choice settings relevant to security, such as data sharing, location and Bluetooth. In particular, users should keep Bluetooth in "off" mode when not intentionally being used to avoid known hacks discussed earlier [9].

**Education on Good Practices.** One of the reasons for the security breaches mentioned in Section 3 is the fact that the service doesn't require the use of strong passwords, leading to higher risk of users repeating the same email-password combination on numerous websites. Weak passwords such as '123456' are still being used by many Fitbit users [12]. It is therefore essential for users to be aware of good practices around building strong passwords for all their accounts.

**Consumer-Friendly Privacy Practices.** Many of the potential risks described above could be partially resolved by providing consumers with much greater control regarding their data and how they choose to use them. For example, users could have the choice to opt out of targeted advertising, having the right to be forgotten regarding all health related and non-health data. In addition, consumers could be asked regularly to update their privacy-friendly defaults [29].

It is therefore essential also for users of wearable devices to be educated about potential privacy and information security related risks to which they are exposed when using these devices.

**Table 1.** Warnings and instructions for using wearable devices

| Functionalities | Warning | Instructions |
|---|---|---|
| Electronic Wallet | Risks of fraud | • Encrypt critical data elements such as ID, passwords and PIN. |
| Data collected | The device can be vulnerable to risks via the Bluetooth or Wifi connection used | • Use an antivirus program to scan any connected device for viruses and malware and determine the settings to limit the personal data captured by the device.<br>• Disable the permission in capturing data such as sleep patterns and capture only the data you really need.<br>• Keep Bluetooth in "off" mode when not intentionally being used. |
| Testing | The device is collecting data such as body temperature, pulse rate, food habits and body weight, steps-distance travelled, calories burned and sleep stage. These data are synchronized to devices that transfer information to Fitbit servers. | • Use a VPN service on all your devices to ensure your privacy.<br>• Think about the purpose and the environment you might want to wear your fitness tracker.<br>• Read the company's privacy policy and ensure that reasonable steps are taken to protect it.<br>• Conduct research on data breaches of a specific device or company and prevention measures taken in the occasion of a future attack. |
| Synchronising with phone or laptop | After the user connects a wearable which has been infected with malware to a PC or laptop, the malware can spread to via the PC or laptop and even infect the entire network. | • Keep your devices and software current with the latest security updates, and the best choice settings relevant to security, such as data sharing, location and Bluetooth. |
| Connecting the device to email | Using email addresses and passwords from third-party websites to log in to Fitbit accounts is risky. | • Use strong passwords and avoid email-password combination repetitions on multiple sites. |

### 5.2 Cybersecurity Guidelines for Manufacturers

A number of guidelines which manufacturers can follow to ensure secure wearable devices are presented below:

**Create Policies and Standards.** Rigorous quality testing should be a standard practice followed by all IoT device manufacturers. Implementing security techniques like strong password protection for security agnostic IoT devices could incur memory and cost overheads, although these practices are essential in securing the network [30]. Organisations can deter some cyber threats by creating appropriate use policies and user agreements for Fitbits and other wearables.

In addition, manufacturers can establish a functionality of the devices informing the user of potential risks to their personal information, when they log into the device application for the first time. Only after the user accepts the warning advice they can log into the device. This way manufacturers can promote general awareness for consumers and support a cybersecurity mind-set for users. By understanding how hackers think, users can take measures to mitigate the potential risks of these devices.

## 6 Discussion

As described above, new IoT devices such as fitness wearable devices, place new threats for users. Lack of security considerations or privacy and data protection considerations around these devices pose serious risks initially to the user of the device with potential cascading effects of a security incident to a large number of devices of the user and others connected to that user. While the data security of fitness wearable devices is questionable [31], a research gap is recognised around security and privacy by design considerations of IoT [32].

Despite these threats, manufacturers of IoT devices still do not provide consumers with enough information about the security features of the devices before they purchase the product. In addition, little information on user behaviour and good practise while using these devices is provided [33].

The General Data Protection Regulation (GDPR) [34] provides users with the rights to their personal data that is held by firms, specifically around the right to be informed of how the data will be used or processed, the right of erasure of data or the right to object the data being collected. Consumers should be given easy to understand security information in order to make correct choices when they shop smart devices. Additionally, users should be given enough information regarding their rights around protecting their personal data after purchasing these devices. However, it would still depend on the consumer and the purchase decision they make based on the level of awareness and knowledge regarding cybersecurity.

Currently, there is lack of awareness of these vulnerabilities by device owners [35], and this is posing difficulties in addressing the security challenges of IoTs [36]. Considering approaches to educate or protect users while imposing security has proved ineffective [35]. Also, creating a culture of fear around potential threats is problematic.

Cybersecurity awareness raising initiatives are necessary in order to increase baseline cybersecurity and develop the skills needed to ensure a safer cyberspace [37]. Users might not realise the risks associated with the use of smart devices [38], this is why it is imperative for all users to gain a basic understanding of the potential harms associated to their use.

The security of IoTs is complex and many stakeholders need to be involved, including the user. There is almost no limit to the variety of devices that can be IoT enabled via wireless connectivity. This is why it is imperative to adopt a proactive IoT-centric security posture, focusing on education and awareness for users.

## 7    Conclusions

The current research has provided a review of the security and privacy related risks related to the use of wearable devices. Users might lack knowledge and awareness around these risks, therefore more efforts are needed from manufacturers of IoT devices to provide consumers with information on user behaviour and good practise while using these devices. Although the focus of this study has been fitness devices, in future work further research will be conducted comparing the risks among different wearable devices.